\documentclass[aps,prl,twocolumn,showpacs,superscriptaddress]{revtex4}
\usepackage{psfrag}
\usepackage{graphicx}
\usepackage{amsfonts}
\usepackage[figuresright]{rotating}
\usepackage{amssymb}
\usepackage{amsmath}
\usepackage{subfigure}
\usepackage{multirow}
\usepackage{tabularx}

\def\Br{\mathbf{r}}

\begin{document}
\title{Topological phases of dipolar particles in elongated Wannier orbitals}
\author{Kai Sun}
\affiliation{Department of Physics, University of Illinois at
  Urbana-Champaign, 1110 West Green Street,  Urbana, IL 61801}
\affiliation{Joint Quantum Institute and Condensed Matter Theory Center, 
Department of Physics, University of Maryland, College Park, MD 20742} 
\author{Erhai Zhao}
\affiliation{Department of Physics and Astronomy, George Mason University, Fairfax, VA 22030}
\affiliation{Department of Physics and Astronomy, University of
Pittsburgh, Pittsburgh, PA 15260}
\author{W. Vincent Liu}
\affiliation{Department of Physics and Astronomy, University of
Pittsburgh, Pittsburgh, PA 15260}
\affiliation{Center for Cold Atom Physics, Chinese Academy of
Sciences, Wuhan 430071, China}

\begin{abstract}
We show that topological phases with fractional excitations can occur 
in two-dimensional ultracold dipolar gases on a particular class of
optical lattices. Due to the dipolar interaction and lattice
confinement, a quantum dimer model emerges naturally as the effective
theory describing the low-energy behaviors of these systems under
well-controlled approximations.
The desired hierarchy of interaction energy scales is achieved by
controlling the anisotropy of the orbital dimers and the dipole
moments of particles. Experimental realization and detection of
various phases are discussed, as well as the possible relevance for
quantum computation.
\end{abstract}
\date{\today}
\maketitle

Unlike the ordered phase described by the Landau
theory of phase transitions, a topological state is characterized by
its response to the changing of  topology of the underlying
manifold, instead of a local order parameter associated with symmetry
breaking~\cite{wen1991}.  Theoretically, these phases are described by
topological field theories, in which the low-energy elementary
excitations often carry fractional quantum numbers of the constituent
particles, a phenomenon known as fractionalization. 
These fractional excitations offer a concrete setting from which new physics
emerges, e.g., the realization of topological quantum computation (
\cite{Nayak2008} and reference therein).

The most celebrated topological phases are perhaps the fractional
quantum Hall (FQH) states described by the topological Chern-Simons gauge
theory, whose low-energy excitations
carry fractional charges and statistics of an electron. Another important example is
the $Z_2$ topological phase in two-dimensional (2D) quantum dimer
models (QDM) of frustrated spins, introduced by Rokhsar and Kivelson (RK)
\cite{Kivelson1987,Rokhsar1988,Moessner2001}. This novel phase is known as the
resonating-valence-bond (RVB) state described by a topological Ising gauge theory and
gives rise to deconfined fractional excitations such as spinons which
carry spin $1/2$ but zero charge \cite{Read1991,Mudry1994,Kitaev2003,Senthil2000,Moessner2001a}.

In ultracold atomic/molecular systems, efforts have been made in
realizing various topological phases. For example, states similar to
the FQH states were proposed in rotating atomic
gases~\cite{Cooper2008}. There also have  been suggestions to simulate
QDM~\cite{blatter2008} and the RVB phase with ultracold atomic
systems, but with the challenge of achieving the
special nonlocal interaction as pointed out~\cite{blatter2008}.
In fact, in contrast to its theoretical appeal, similar difficulties 
appear in most attempts to realize QDM and the RVB phase, for example, 
in the frustrated-Ising model \cite{Moessner2000,Moessner2001b} and the easy-axis antiferromagnet \cite{Balents2002,Isakov2006}. 

In this letter, we propose a different approach to realize and study
the topological phases using ultracold dipolar particles (atoms or
molecules) in specifically designed, non-standard optical lattices, in
part motivated by the recent experiments of polar
molecules (\cite{Ni+Jin+Ye:08,Ospelkaus-2008,Lahaye+Pfau:08pre_rev}
and references therein).
We exploit two novel features: 1) the orbital degree of freedom
controlled by optical
lattice confinement, and 2) the long-range interaction associated with
dipoles. Using a site-to-bond duality mapping \cite{dual}, we show that the orbital
motions on some optical lattices with dipolar interactions can be
mapped 
onto a QDM: a particle in the elongated Wannier orbital
on the optical lattice corresponds to a
dimer on the dual lattice; and the
dipolar interaction between particles
enforces the ``hard-core constraint'' of the
dimers. By tuning the optical lattice and the external field
(the latter
aligns the dipole moments), this system shows great tunablity in
exploring different phases of the QDM. For bosonic particles, within
experimentally reasonable parameter ranges, we shall show that various
phases can be stabilized, including conventional crystalline
phases and a $Z_2$ topological phase. The latter is
similar to the RVB phase of the
QDM, and will be referred to as the orbital-RVB (ORVB).
Experimental detection of these phases and the fractional
excitations in the ORVB phase shall be discussed. More interestingly,
the orbital dimers in our model system
can be either bosonic or fermionic, depending on the statistics of the constituent
particles, whereas the frustrated spin
system only allows bosonic dimers. This opens a new possibility for studying fermionic QDM
and even the mixture of dimers with different statistics.

\begin{figure}
\includegraphics[width=3 in]{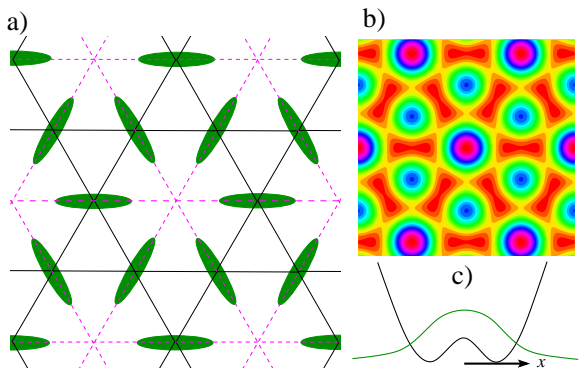}
\caption{(Color online) a) The Kagome lattice (solid lines) and its dual triangular lattice (dash lines).
The green ellipse represents the elongated wavefunction at each site of the Kagome lattice (the bond of the triangular lattice).
b) Contour plot of the 2D double-well optical lattice potential (with $I_2=2 I_1$). Dipolar atoms (or molecules) reside
at the intensity minima, which form the Kagome lattice. c) The local double-well potential and the wavefunction, which can be approximated by a Guassian.}
\label{fig:kagome}
\end{figure}

\textit{Orbital dimers in 2D optical lattices.}
We focus on a concrete example, a 2D Kagome lattice constituted by anisotropic potential wells
[Fig. \ref{fig:kagome}(b)]. Due to this anisotropy, the Wannier wavefunction at each site would be
elongated and would be considered as a ``dimer'' as shown below.
Note that arbitrary lattice potential can be generated in 2D for example by projecting 
a holographic mask through an imaging system \cite{markus}. Such potential can also be 
achieved by independent sets of laser beams to produce an intensity pattern of
(similar to Ref.~\cite{Santos+Lewenstein:04})
\begin{align}
I=I_1 \sum_{i=1,2,3}\cos^2(\mathbf{k}_i\cdot\mathbf{r})
+ I_2 \sum_{j=1,2,3}\cos^2 (\mathbf{q}_j\cdot\mathbf{r}),
\end{align}
where the amplitude of the wavevectors projected on the plane satisfy
$|\mathbf{q}_j|=\sqrt{3}|\mathbf{k}_i|$ (the $z$ component is
irrelevant in a 2D system) and the azimuth angle of
$\mathbf{k}_i$s ($\mathbf{q}_j$s) are $\theta_{\mathbf{k}_i}=i
\pi/3$ ($\theta_{\mathbf{q}_j}=(j+1/2) \pi/3$).
This intensity pattern induces an optical lattice consisting of
double-well-like potentials [Fig. \ref{fig:kagome}(c)]. However, since 
the energy barrier between the two potential wells at one site is small, the double-well structure
can be treated as a single anisotropic potential well, unless the energy of the trapped state is 
much smaller than the recoil energy, which is the limit we would avoid in this letter.

In fact, for our interest, the precise shape of the potential is of little
importance, as long as the ground state wavefunction at each site is
{\it sufficiently elongated} and pointing towards the center of each
hexagon.
We call this elongated wavefunction an ``orbital dimer'',
because its spatial symmetry mimics that of a link variable defined on
bonds. To make this analogy precise, we perform a site-to-bond duality
transformation. As shown in Fig.~\ref{fig:kagome}(a), under this
mapping, an atom/molecule localized at a site of the original Kagome
lattice (solid lines) becomes a dimer residing on a bond of the dual
triangular lattice (dashed lines). Accordingly, particles 
on the original lattice are mapped to dimers on the dual lattice~\cite{mapping}.
\nocite{Freedman2005,discussion}

\begin{figure}
\begin{center}
\subfigure[$V_1$]{\includegraphics[width=0.09\textwidth]{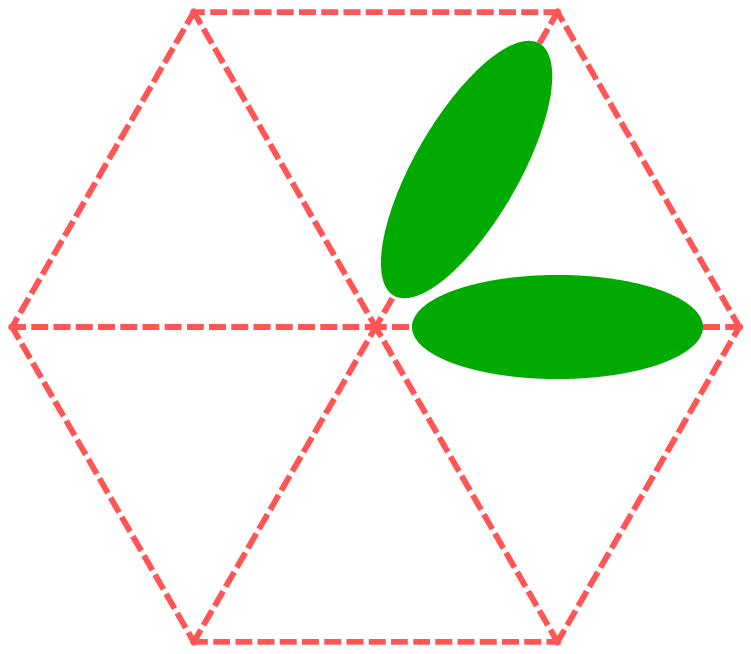}\label{fig:v1}}
\subfigure[$V_2$]{\includegraphics[width=0.09\textwidth]{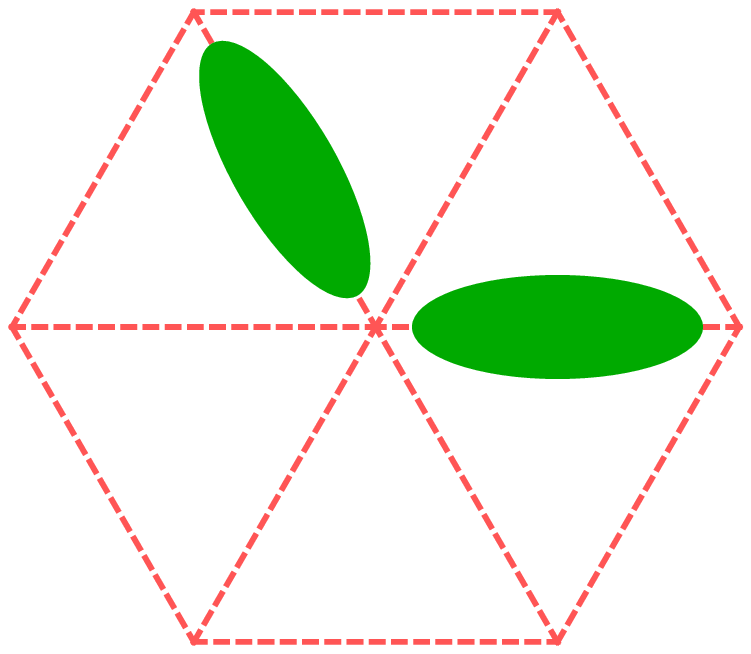}\label{fig:v2}}
\subfigure[$V_3$]{\includegraphics[width=0.09\textwidth]{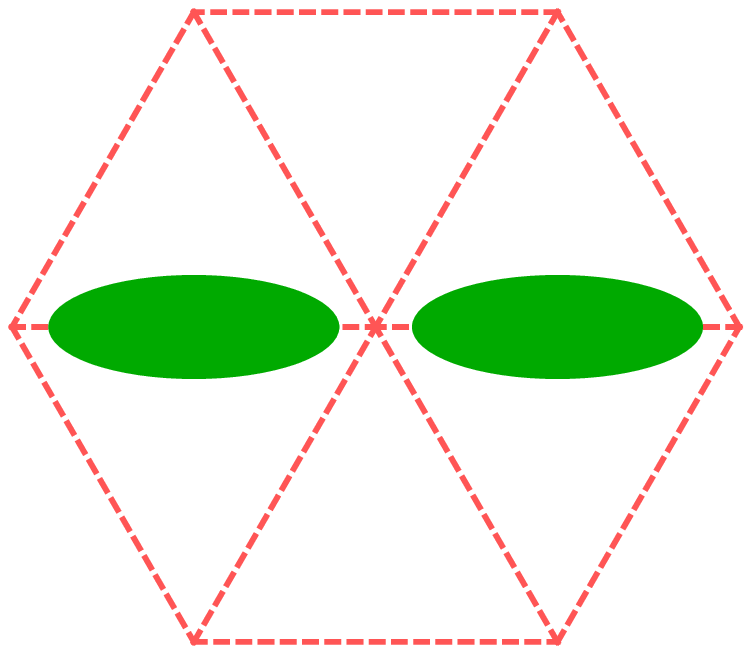}\label{fig:v3}}
\subfigure[$V_\parallel$]{\includegraphics[width=0.09\textwidth]{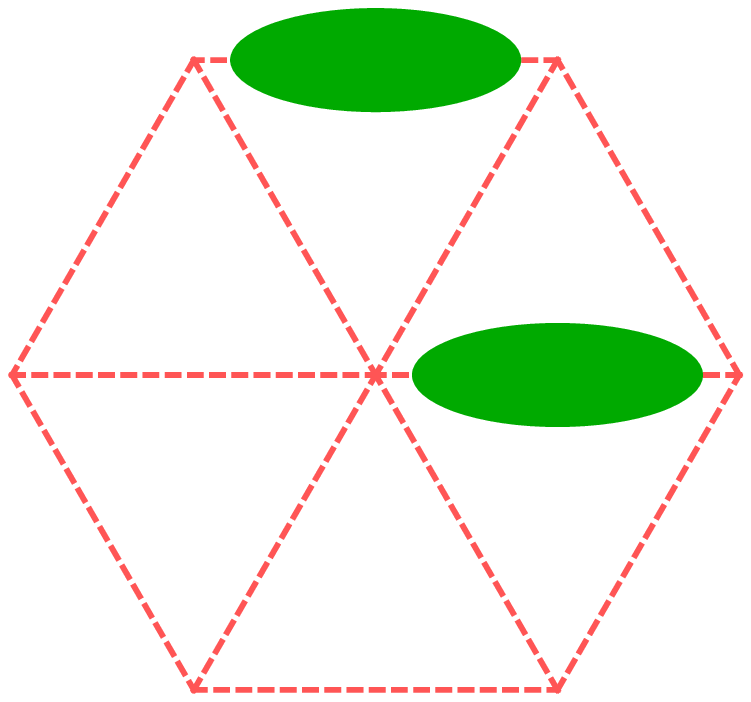}\label{fig:v_parallel}}
\subfigure[$V_\textrm{S}$]{\includegraphics[width=0.09\textwidth]{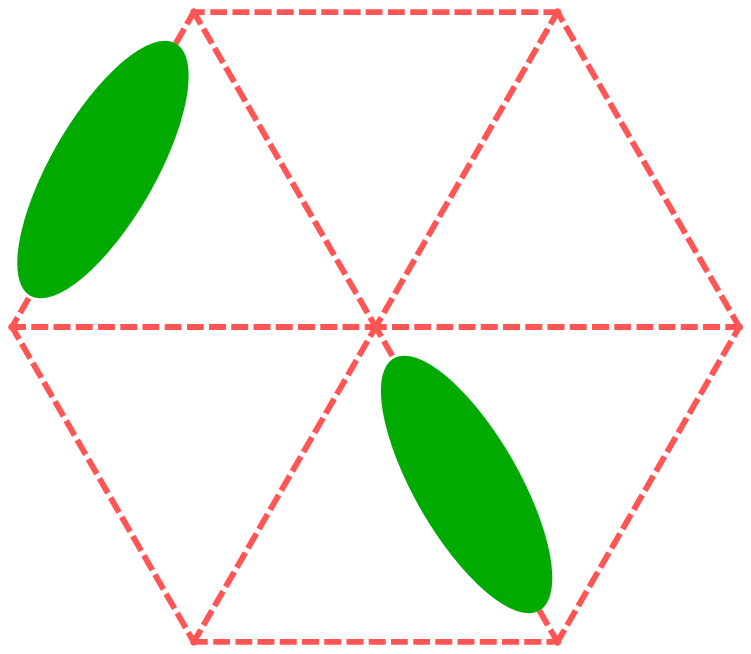}\label{fig:v_stagger}}
\end{center}
\caption{(Color online) Interactions between dipolar particles on the  dual triangular lattice shown in Fig. \ref{fig:kagome}.}
\label{fig:interactions}
\end{figure}

{\it Dipolar interactions.}
If the dipolar particles are polarized in the $z$-direction by an external field, their interaction
is given by the long-range potential $V_{dd}(\Br)=d^2(1-3 z^2/r^2)/r^3$. Here,
$d$ is the dipole moment, $r$ is the distance between the two dipoles,
and $z$ is the $z$-component of $\Br$. For a quasi-$2D$ system, the dipolar interaction is
repulsive near the $xy$-plane, which leads to energy penalties for particles
close to each other. The largest interaction is the on-site repulsion $U$ and the next one
is the repulsion between particles on the same hexagon of the original lattice, or equivalently
dimers joining at the same site on the dual lattice. Three such configurations $i=1,2,3$ are shown in
Fig.~\ref{fig:interactions}(a)-(c). Even smaller is the interaction for parallel ($V_\parallel$) and staggered ($V_\textrm{S}$) dimers, shown in Figs.~\ref{fig:v_parallel} and \ref{fig:v_stagger}, respectively.
The effect of longer-range interactions will be studied below.
In the limit $U > V_i \gg V_\parallel > V_\textrm{S}$, below a critical filling ($1/6$ for the Kagome lattice), 
the large energy cost $V_i$ ($i=1,2,3$) forbids dimers to share a common site on the dual
lattice. This enforces the ``hard-core constraint", a key feature of QDMs. At the critical filling, the dual lattice is ``fully covered" by dimers. Namely, among the six bonds connected to any site, one and
only one bond is occupied by a dimer.

We now show that the desired hierarchy of interactions given above can be satisfied by 
controlling the dipolar interactions and increasing the elongation of Wannier wavefunctions.
For this purpose, we approximate the Wannier function at each site of the Kagome lattice by these
Gaussian ``atomic orbitals",
\[
w(x,y,z)=(\alpha\beta\gamma)^{-1/2}\pi^{-3/4}\exp [-\frac{x^2}{2\alpha^2}-\frac{y^2}{2\beta^2}-\frac{z^2}{2\gamma^2}],
\]
with $\alpha>\beta\gg \gamma$. Here, a strong laser field in the $z$ direction is used to confine
the particles within the $xy$ plane ($\gamma$ being small) and the $x$ axis is
chosen to be along the direction of the double-well potential as shown
in Fig. \ref{fig:kagome}(c). This simple ansatz would not provide an accurate value for $U$ and $V$s, 
but it suffices to estimate their order of magnitude and demonstrate the proper limit in which our system
could be mapped to a QDM.

Within this ansatz, we find
\[
U\simeq d^2 \sqrt{\frac{2}{\pi}}\left[\frac{2}{3\alpha\beta\gamma}-\frac{E(1-\alpha^2/\beta^2)}{\alpha^2\beta}\right].
\]
where $E(x)$ is the complete elliptic integral of the second kind.
For tight confinement (small $\beta$ and $\gamma$), $U$ is exceedingly
large, so it disfavors multi-occupancy of the same lattice site.
On the dual lattice, this means that there can be at most one dimer on
each bond. It is also straightforward to compute the corresponding interaction $V_i$. For example,
\[
V_3=\frac{d^2}{r^3}\sqrt{\frac{2}{\pi}} \int_0^{\frac{\pi}{2}} d
\theta \sec^3\theta e^{-\frac{y}{2}}
(y^{\frac{5}{2}}-y^{\frac{3}{2}})
+\frac{d^2\sqrt{8\pi}}{3\alpha\beta\gamma}e^{-\frac{r^2}{2\alpha^2}},
\]
where $y(\theta)=r^2/(\alpha^2+\beta^2\tan^2\theta)$, and $r$ is the
distance between the center of two dimers.

Since the lattice structure is unspecified and can vary arbitrarily~\cite{markus}, 
we can treat $\alpha$, $\beta$, and $\gamma$ as free controlling parameters and show 
that the limit $U > V_i \gg V_\parallel > V_\textrm{S}$ can be achieved with 
elongated Wannier functions.
First, it is useful to consider the case of isotropic Wannier functions,
$\alpha=\beta$, and small wavefunction overlap, $\alpha\ll r$  with
$r$ being the distance between two lattice sites.
Then, the interaction energy is reduced to  $d^2/r^3$, and
the ratio $V_1:V_2:V_3:V_\parallel:V_\textrm{S}$ is  $8:1.54:1:1:0.43$.
In particular, we see that $V_2$ and $V_\parallel$ are of the same order.
A key point of our proposal is that by using elongated Wannier functions,
the dipolar interaction for configuration $i=2,3$ can be boosted with respect
to the parallel configuration, since they have much larger wave
function overlap (schematically shown in Fig. \ref{fig:interactions}).
For example, for $\alpha=0.3$ and $\beta=0.1$ (in the units of the
lattice constant), the ratio $V_1:V_2:V_3:V_\parallel:V_\textrm{S}$ becomes $30:8.6:8.6:1:0.49$
for $\gamma=\beta/10$ and $21:5.8:5.6:1:0.49$ for $\gamma=\beta/5$.
Here $V_3$ is almost one order of magnitude larger than $V_{\parallel}$ and their 
ratio can be further enhanced by increasing the elongation of the Wannier function 
(through adjusting the optical lattice) and/or reducing $\gamma$. Notice that $\gamma$ 
can be tuned independently. For a lattice with sufficient elongated Wannier functions, 
the mapping to QMD becomes asymptotically accurate as $\gamma$ is reduced.

\textit{The orbital dimer model.}
At the critical filling, the low-energy fluctuations of the system are described by the following Hamiltonian after
projecting out high-energy degrees of freedom (at the scale of $U$ and $V_i$),
\begin{align}
H=
&-t_{\textrm{eff}}
\sum \left(|\setlength{\unitlength}{3158sp}%
\begingroup\makeatletter\ifx\SetFigFont\undefined%
\gdef\SetFigFont#1#2#3#4#5{%
  \reset@font\fontsize{#1}{#2pt}%
  \fontfamily{#3}\fontseries{#4}\fontshape{#5}%
  \selectfont}%
\fi\endgroup%
\begin{picture}(319,210)(517,-186)
\thinlines
\multiput(720,-10)(1.54799,-2.57998){58}{\makebox(2.0833,14.5833){\SetFigFont{5}{6}{\rmdefault}{\mddefault}{\updefault}.}}
\multiput(718,-10)(-1.55263,-2.58772){58}{\makebox(2.0833,14.5833){\SetFigFont{5}{6}{\rmdefault}{\mddefault}{\updefault}.}}
\multiput(548,-10)(1.55263,-2.58772){58}{\makebox(2.0833,14.5833){\SetFigFont{5}{6}{\rmdefault}{\mddefault}{\updefault}.}}
\multiput(718,-10)(-3.00000,0.00000){58}{\makebox(2.0833,14.5833){\SetFigFont{5}{6}{\rmdefault}{\mddefault}{\updefault}.}}
\multiput(805,-158)(-3.00000,0.00000){58}{\makebox(2.0833,14.5833){\SetFigFont{5}{6}{\rmdefault}{\mddefault}{\updefault}.}}
\put(803,-153){\line(-1, 0){171}}
\put(722, -9){\line(-1, 0){172}}
\end{picture}
\rangle
\langle \setlength{\unitlength}{3158sp}%
\begingroup\makeatletter\ifx\SetFigFont\undefined%
\gdef\SetFigFont#1#2#3#4#5{%
  \reset@font\fontsize{#1}{#2pt}%
  \fontfamily{#3}\fontseries{#4}\fontshape{#5}%
  \selectfont}%
\fi\endgroup%
\begin{picture}(317,216)(525,-421)
\thinlines
\multiput(720,-235)(1.54799,-2.57998){58}{\makebox(2.0833,14.5833){\SetFigFont{5}{6}{\rmdefault}{\mddefault}{\updefault}.}}
\multiput(718,-235)(-1.55263,-2.58772){58}{\makebox(2.0833,14.5833){\SetFigFont{5}{6}{\rmdefault}{\mddefault}{\updefault}.}}
\multiput(548,-235)(1.55263,-2.58772){58}{\makebox(2.0833,14.5833){\SetFigFont{5}{6}{\rmdefault}{\mddefault}{\updefault}.}}
\multiput(718,-235)(-3.00000,0.00000){58}{\makebox(2.0833,14.5833){\SetFigFont{5}{6}{\rmdefault}{\mddefault}{\updefault}.}}
\multiput(805,-383)(-3.00000,0.00000){58}{\makebox(2.0833,14.5833){\SetFigFont{5}{6}{\rmdefault}{\mddefault}{\updefault}.}}
\thicklines
\multiput(809,-386)(-5.48713,9.14522){17}{\makebox(8.3333,12.5000){\SetFigFont{7}{8.4}{\rmdefault}{\mddefault}{\updefault}.}}
\multiput(644,-388)(-5.53125,9.21875){17}{\makebox(8.3333,12.5000){\SetFigFont{7}{8.4}{\rmdefault}{\mddefault}{\updefault}.}}
\end{picture}
|+h.c. \right) \nonumber\\
&+ V_{\textrm{eff}}
\sum \left( |\setlength{\unitlength}{3158sp}%
\begingroup\makeatletter\ifx\SetFigFont\undefined%
\gdef\SetFigFont#1#2#3#4#5{%
  \reset@font\fontsize{#1}{#2pt}%
  \fontfamily{#3}\fontseries{#4}\fontshape{#5}%
  \selectfont}%
\fi\endgroup%
\begin{picture}(319,210)(517,-186)
\thinlines
\multiput(720,-10)(1.54799,-2.57998){58}{\makebox(2.0833,14.5833){\SetFigFont{5}{6}{\rmdefault}{\mddefault}{\updefault}.}}
\multiput(718,-10)(-1.55263,-2.58772){58}{\makebox(2.0833,14.5833){\SetFigFont{5}{6}{\rmdefault}{\mddefault}{\updefault}.}}
\multiput(548,-10)(1.55263,-2.58772){58}{\makebox(2.0833,14.5833){\SetFigFont{5}{6}{\rmdefault}{\mddefault}{\updefault}.}}
\multiput(718,-10)(-3.00000,0.00000){58}{\makebox(2.0833,14.5833){\SetFigFont{5}{6}{\rmdefault}{\mddefault}{\updefault}.}}
\multiput(805,-158)(-3.00000,0.00000){58}{\makebox(2.0833,14.5833){\SetFigFont{5}{6}{\rmdefault}{\mddefault}{\updefault}.}}
\put(803,-153){\line(-1, 0){171}}
\put(722, -9){\line(-1, 0){172}}
\end{picture}
\rangle
\langle \setlength{\unitlength}{3158sp}%
\begingroup\makeatletter\ifx\SetFigFont\undefined%
\gdef\SetFigFont#1#2#3#4#5{%
  \reset@font\fontsize{#1}{#2pt}%
  \fontfamily{#3}\fontseries{#4}\fontshape{#5}%
  \selectfont}%
\fi\endgroup%
\begin{picture}(319,210)(517,-186)
\thinlines
\multiput(720,-10)(1.54799,-2.57998){58}{\makebox(2.0833,14.5833){\SetFigFont{5}{6}{\rmdefault}{\mddefault}{\updefault}.}}
\multiput(718,-10)(-1.55263,-2.58772){58}{\makebox(2.0833,14.5833){\SetFigFont{5}{6}{\rmdefault}{\mddefault}{\updefault}.}}
\multiput(548,-10)(1.55263,-2.58772){58}{\makebox(2.0833,14.5833){\SetFigFont{5}{6}{\rmdefault}{\mddefault}{\updefault}.}}
\multiput(718,-10)(-3.00000,0.00000){58}{\makebox(2.0833,14.5833){\SetFigFont{5}{6}{\rmdefault}{\mddefault}{\updefault}.}}
\multiput(805,-158)(-3.00000,0.00000){58}{\makebox(2.0833,14.5833){\SetFigFont{5}{6}{\rmdefault}{\mddefault}{\updefault}.}}
\put(803,-153){\line(-1, 0){171}}
\put(722, -9){\line(-1, 0){172}}
\end{picture}
|+
|\setlength{\unitlength}{3158sp}%
\begingroup\makeatletter\ifx\SetFigFont\undefined%
\gdef\SetFigFont#1#2#3#4#5{%
  \reset@font\fontsize{#1}{#2pt}%
  \fontfamily{#3}\fontseries{#4}\fontshape{#5}%
  \selectfont}%
\fi\endgroup%
\begin{picture}(317,216)(525,-421)
\thinlines
\multiput(720,-235)(1.54799,-2.57998){58}{\makebox(2.0833,14.5833){\SetFigFont{5}{6}{\rmdefault}{\mddefault}{\updefault}.}}
\multiput(718,-235)(-1.55263,-2.58772){58}{\makebox(2.0833,14.5833){\SetFigFont{5}{6}{\rmdefault}{\mddefault}{\updefault}.}}
\multiput(548,-235)(1.55263,-2.58772){58}{\makebox(2.0833,14.5833){\SetFigFont{5}{6}{\rmdefault}{\mddefault}{\updefault}.}}
\multiput(718,-235)(-3.00000,0.00000){58}{\makebox(2.0833,14.5833){\SetFigFont{5}{6}{\rmdefault}{\mddefault}{\updefault}.}}
\multiput(805,-383)(-3.00000,0.00000){58}{\makebox(2.0833,14.5833){\SetFigFont{5}{6}{\rmdefault}{\mddefault}{\updefault}.}}
\thicklines
\multiput(809,-386)(-5.48713,9.14522){17}{\makebox(8.3333,12.5000){\SetFigFont{7}{8.4}{\rmdefault}{\mddefault}{\updefault}.}}
\multiput(644,-388)(-5.53125,9.21875){17}{\makebox(8.3333,12.5000){\SetFigFont{7}{8.4}{\rmdefault}{\mddefault}{\updefault}.}}
\end{picture}
\rangle
\langle \setlength{\unitlength}{3158sp}%
\begingroup\makeatletter\ifx\SetFigFont\undefined%
\gdef\SetFigFont#1#2#3#4#5{%
  \reset@font\fontsize{#1}{#2pt}%
  \fontfamily{#3}\fontseries{#4}\fontshape{#5}%
  \selectfont}%
\fi\endgroup%
\begin{picture}(317,216)(525,-421)
\thinlines
\multiput(720,-235)(1.54799,-2.57998){58}{\makebox(2.0833,14.5833){\SetFigFont{5}{6}{\rmdefault}{\mddefault}{\updefault}.}}
\multiput(718,-235)(-1.55263,-2.58772){58}{\makebox(2.0833,14.5833){\SetFigFont{5}{6}{\rmdefault}{\mddefault}{\updefault}.}}
\multiput(548,-235)(1.55263,-2.58772){58}{\makebox(2.0833,14.5833){\SetFigFont{5}{6}{\rmdefault}{\mddefault}{\updefault}.}}
\multiput(718,-235)(-3.00000,0.00000){58}{\makebox(2.0833,14.5833){\SetFigFont{5}{6}{\rmdefault}{\mddefault}{\updefault}.}}
\multiput(805,-383)(-3.00000,0.00000){58}{\makebox(2.0833,14.5833){\SetFigFont{5}{6}{\rmdefault}{\mddefault}{\updefault}.}}
\thicklines
\multiput(809,-386)(-5.48713,9.14522){17}{\makebox(8.3333,12.5000){\SetFigFont{7}{8.4}{\rmdefault}{\mddefault}{\updefault}.}}
\multiput(644,-388)(-5.53125,9.21875){17}{\makebox(8.3333,12.5000){\SetFigFont{7}{8.4}{\rmdefault}{\mddefault}{\updefault}.}}
\end{picture}
|\right)+V_{LR},
\label{eq:H}
\end{align}
where the summation is over all flippable plaquettes~\cite{Moessner2001} 
and $V_{LR}$ describes higher order terms due to longer-range interactions
(beyond those shown in Fig. \ref{fig:interactions}).
Here the dimer has the same statistical property as
the constituent particle, so it could be either \emph{bosonic} or \emph{fermionic}.
At the critical filling, the dual triangular lattice is fully covered by the orbital dimers. 
All local density fluctuations are gapped out due to large energy penalty of $V_i$ 
($i=1,2,3$). Hence density remains homogeneous on large length scales
and the dominant effect of longer-range repulsive interaction is to contribute a constant shift 
in the free energy. As a leading order approximation, we will drop $V_{LR}$ in the 
low-energy effective Hamiltonian Eq. \eqref{eq:H} and expect the orbital dimer model 
captures the essential low-energy physics of the system.

At the critical filling, particle hopping gives rise to overlapping
dimers (sharing the same site), which costs
energy of order $V_i$. Such virtual hopping processes lead to
quantum resonance of orbital dimers around a plaquette ($t_\textrm{eff}$),
as well as corrections to $V_{\parallel}$ and $V_{\textrm{S}}$
through second order perturbations, of order $t^2 /{V_i}$.
To the leading approximation, we use one (average) energy scale $t$ to estimate the
amplitudes of relevant hoppings.
The repulsive interaction between two
parallel dimers, $V_{\textrm{eff}}$, is set by the energy scale
$V_\parallel-V_\textrm{S}$ up to corrections of $O(t^2/V_i)$.
By varying $t$ and the interacting strength,
$t_{\textrm{eff}}$ and $V_{\textrm{eff}}$  can be tuned to explore various
parameter ranges of the QDM, except for the limit of $V_{\textrm{eff}} \ll -t^2/{V_i}$,
since $V_\parallel>V_\textrm{S}$ for dipolar interaction.
To estimate the relevant energy scales, we
take $d=0.1$ Debye, $a=0.25\mu$m, $V_{1}\sim 4 V_{2,3}\sim
20V_\parallel$, and $t\sim 10$nK. Then, $V_{\textrm{eff}}\sim V_\parallel-V_\textrm{S}\sim 2.3$nK and
$t_{\textrm{eff}}\sim 2t^2/V_1\sim 2.16$nK.
This shows that the
condition $U,V_i \gg V_{\textrm{eff}}\sim t_{\textrm{eff}} \gg T$
can be satisfied. 

\begin{figure}
\begin{center}
\includegraphics[width=3in]{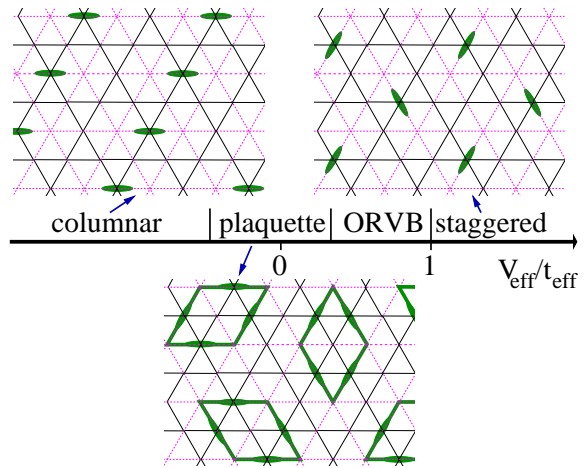}
\end{center}
\caption{(Color online) The phase diagram for the bosonic particles and the particle configuration
in the crystalline phases.
}
\label{fig:phase}
\end{figure}

For bosonic particles, the phase diagram 
at the critical filling is shown in Fig. \ref{fig:phase}, using the knowledge of QDM
\cite{Moessner2001}.
As $V_{\textrm{eff}}/t_{\textrm{eff}}$ increases,
four phases appear consequently: the columnar phase,
the $\sqrt{12}\times\sqrt{12}$ plaquette phase
($V_{\textrm{eff}}/t_{\textrm{eff}}\sim
0$), the ORVB topological phase ($V_{\textrm{eff}}/t_{\textrm{eff}}\lesssim
1$), and the staggered phase ($V_{\textrm{eff}}/t_{\textrm{eff}}>1$).  In the staggered
and the columnar phases, the dimers are frozen in the ground
state, while in the $\sqrt{12}\times\sqrt{12}$ state, they can
resonate in each plaquette marked by the thick lines in
Fig.~\ref{fig:phase}. In terms of dipolar particles,
all these three phases correspond to various
commensurate charge-density-wave states, which break the point group
symmetry of the underlying lattice. On the other hand, all the spatial
symmetries of the lattice are preserved in the ORVB phase. 
However, this state has critical dependence
on the topology of the underlying manifold. On a topologically
nontrivial manifold, it shows degenerate ground states, 
topologically distinct from one another,
known as topological degeneracy.  As a $Z_2$
topological phase, the ORVB phase has $4$ degenerate ground states
distinguished by two $Z_2$ topological indices if placed 
on a torus.  As discussed below,
the phenomenon of topological degeneracy not
only provides unique experimental signatures for the ORVB phase, but
also leads to the possibility of realizing quantum memory.

In the context of frustrated spins
the RVB state has two different types of fractional excitations: spinons (carrying spin 1/2)
and holons (carrying charge $e$), known as the spin-charge
separation. For an ORVB state, however, the orbital dimers
are not spin singlets, and hence spinon excitations are forbidden. As
a result, only one type of fractional excitations, holons, emerges.  A
holon carries only half the quantum number of an underlying dipolar
particle, e.g., a half dipole, while the statistics of a holon may depend on
microscopic details of the system \cite{Kivelson1989}. This phenomenon
shares strong similarities with the deconfined monopole excitations in
spin ice \cite{Castelnovo2008}. 

Using different optical lattices of double-well potentials, QDM on
other lattices can also be achieved through similar constructions. For
example, the following laser intensity results in a square-lattice QDM,
\begin{align}
I=I_1 \sum_{i=1,2}\cos ^2(\mathbf{k}_i\cdot\mathbf{r})
+ I_2 \sum_{j=1,2}\cos^2 (\mathbf{q}_j\cdot\mathbf{r}).
\end{align}
with the in-plane component satisfying $|\mathbf{q}_j|=3|\mathbf{k}_i|/\sqrt{2}$
and the azimuth angles: $\theta_{\mathbf{k}_i}=(2 i+1)\pi/4$ and $\theta_{\mathbf{q}_j}=2 j \pi/4$.
In this case, the ORVB phase shrinks into a RK point as expected for all
bipartite lattices, while the crystalline phases remain to occur in
regions with dimer configurations  slightly different from the triangular lattice.

\textit{Experimental detection.}
Each of the phases discussed above
displays a distinct interference pattern in Bragg
scattering experiments. In the lattice shown in Fig.~\ref{fig:kagome},
for example, the ORVB phase has Bragg peaks at wavevectors
commensurate with the reciprocal vectors of the lattice: $(2\pi,2
\pi/\sqrt{3})$ and $(0, 4\pi/\sqrt{3})$.
In the staggered phase, however, the Bragg peaks are
located at $(m \pi, 2 n\pi/\sqrt{3})$ for any integer $m$ and
$n$. Similarly, the columnar phase [Bragg peaks commensurate with
vectors $(\pi,\sqrt{3}\pi)$ and $(0,4\pi/\sqrt{3})$] and the
$\sqrt{12}\times\sqrt{12}$ phase [with vectors
$(\pi/3,\pm\pi/\sqrt{3})$] can also be distinguished.
The six-fold rotational symmetry is broken in the 
staggered and columnar phase. This can be used to distinguished them 
from other phases in time-of-flight experiments.

As in the experimental search for spin liquids, lack of symmetry breaking 
at low temperatures is an evidence of the ORVB phase in our system. 
One way to distinguish it from the topologically trivial state is to detect the topological
degeneracy in, for example, the ``annulus'' geometry, which can be achieved by punching a hole
on the 2D optical lattice using an extra laser beam to form a forbidden region with high energy barrier.
For a large enough hole, the ORVB ground state is doubly
degenerate. The two topologically-distinct 
degenerate states can
serve as a ``qubit'' \cite{Moessner2002}. It can be
``flipped'' by creating a pair of holons and then annihilating them after
moving one holon around the hole of the annulus. Furthermore,
topological quantum memory can be realized by punching multiple holes and
manipulating the holons.
For a hole of finite size, 
tunneling lifts the degeneracy and results in two nearly-degenerate
states. The energy splitting increases upon reducing the size of the hole.
The size dependence of the energy splitting can serve as a
direct evidence of $Z_2$ topological order.
However, it is challenging to measure the splitting, which is on the order of $pK$, experimentally.
Another distinct feature of the ORVB state is the holons.
How to efficiently probe such fractional excitations remains an important open problem.

\begin{acknowledgments}
We thank Sankar Das Sarma, Eduardo Fradkin, Han Pu, Hong Yao, and especially Christopher
Henley for helpful discussions. This work is supported by
the Office of Science, U.S. Department of Energy under Contracts
DE-FG02-91ER45439 of the Frederick Seitz Materials Research Laboratory
at the University of Illinois, JQI-NSF-PFC and JQI-AFOSR-MURI at Maryland (KS),
U.S. Army Research Office Grant No. W911NF-07-1-0293 (EZ and WVL), 
NIST Grant 70NANB7H6138 Am 001 and ONR Grant N00014-09-1-1025A (EZ),
and the
CAS/SAFEA International Partnership Program for Creative Research
Teams of China (WVL).
\end{acknowledgments}


\begin{thebibliography}{29}
\bibitem[{dua()}]{dual}
\bibinfo{note}{In this letter, a duality relation is always referred to as this
  site-to-bond mapping.}
\end{thebibliography}

\end{document}